\documentclass[12pt]{iopart}
\newcommand{\sss}{\scriptstyle}
\begin{document}
\title[On the mutual polarization of  two $He^4$ atoms]{On the mutual polarization of two $He^4$ atoms}

\author{Vadim M Loktev$^1$ and Maksim D Tomchenko$^2$}

\address{Bogolyubov Institute for Theoretical Physics, 14b,
 Metrolohichna Str., 03680 Kiev, Ukraine}
\ead{vloktev@bitp.kiev.ua; mtomchenko@bitp.kiev.ua}

\begin{abstract}
We propose a simple method based on the standard quantum-mechanical perturbation theory to calculate the mutual polarization of two atoms $He^4$.
\end{abstract}
\pacs{32.10.Dk}
\submitto{Paper}
\maketitle
\normalsize

   In works \cite{wb1,wb2}, it was  shown
          that each of two interacting $He^4$ atoms induces a dipole moment
     \begin{equation}
  \bi{d} = -D_{7} |e| \frac{a_{\rm{B}}^{8}}{R^7} \bi{n}, \quad D_{7}=13.2 \mbox{$-$}
  18.4
  \label{2-0}     \end{equation}
     on the neighboring atom. Here, $\bi{n} =\bi{R}/R$ is the unit vector toward the second atom, and
     $a_{\rm{B}} = \hbar^2/me^2=0.529\,\mbox{\AA}$. We will call the polarization
mechanism ``tidal'', since the deformation of electron shells
  of atoms in this case reminds gravitational tides. The calculation in \cite{wb1,wb2}
     was executed on the basis of special dispersion relations known mainly to experts in atomic physics.
      Therefore, we propose to solve the problem by a simpler method which does not leave the frame of the ordinary
      course of quantum mechanics and, in this case, will verify
     the result  (\ref{2-0}).
     In addition, the presented calculations are useful \cite{pov1} for the explanation of the recently discovered electric activity of superfluid He II \cite{r1,r2}.

           According to the standard perturbation theory (PT), we will search for the ground-state wave function for
     two immovable interacting atoms
    $He^4$, A and B, as the expansion in all possible states
    $\Psi_{\tilde{m}}$ of two free atoms A and B:
        \begin{equation}
  \Psi^{\rm{AB}}_0 = \sum\limits_{\tilde{m}_{\rm{A}}\tilde{m}_{\rm{B}}}c_{\tilde{m}_{\rm{A}}\tilde{m}_{\rm{B}}}
   \Psi^{\rm{A}}_{\tilde{m}_{\rm{A}}}\Psi^{\rm{B}}_{\tilde{m}_{\rm{B}}},
      \label{2-1}     \end{equation}
        where  $\tilde{m}_{\rm{A}}$ and $\tilde{m}_{\rm{B}}$  are the complete collections of quantum numbers $(n, l, m)$
      for these atoms. By $\Psi_{\tilde{l}}\equiv |\tilde{l}\rangle $, we denote the
      set of states with a given $l$ and all possible corresponding  $m$ and  $n$  ($m=0, \pm 1, ... \pm l$, $\ n=l,...,\infty$).
       In what follows, we will  omit sometimes the tilde for simplicity: $\Psi_{\tilde{l}}\equiv \Psi_{l}$,
   $ |\tilde{l}\rangle \equiv |l\rangle $,  $c_{\tilde{0}\tilde{0}} \equiv c_{00}$,
     \textit{etc}.  Since the excited s-states  ($1\rm{s}\it{n}\rm{s}$) give the zero contribution to a DM
        (\ref{2-3}), we neglect them, so that $\Psi_{0}$ means the ground state $1\rm{s}1\rm{s}$ of an atom $He^4$.
         It is obvious that $c_{ij}=c_{ji}$. According to PT, we have
       \begin{equation}
    c_{ij}=c_{ij}^{(1)}+c_{ij}^{(2)}+\ldots,
\label{2-2}     \end{equation}
      where the upper index shows the order of PT. The tidal DM of atom A is sought by the general formula
    \begin{equation}
  \bi{d}^{\rm{A}} = \int \Psi^{*\rm{AB}}_0 \hat{\bi{d}}^{\rm{A}}
   \Psi^{\rm{AB}}_0 \rmd\bi{r}_1^{\rm{A}} \rmd\bi{r}_2^{\rm{A}} \rmd\bi{r}_1^{\rm{B}}
   \rmd\bi{r}_2^{\rm{B}},
    \label{2-3}     \end{equation}
    where $\bi{r}_j$ are the coordinates of electrons, and
    \begin{equation}
     \hat{ \bi{d}}^{\rm{A}}= e(\bi{r}_1^{\rm{A}} + \bi{r}_2^{\rm{A}}).
     \label{2-3b}     \end{equation}
     The function $\Psi_0$ describes a parastate. Therefore, the coefficients
     $c_{ij}$ are nonzero only for parastates $\Psi_j$.
     In view of this fact, the spin function is the same for all $\Psi_j$ in
    (\ref{2-1}). Therefore, the spins can be took out in (\ref{2-3}). Since the sum over spins gives 1, we omit the
spin coordinates.

       We note that the quantity $d^{\rm{A}} \sim c_{10}(1) \sim (a_{\rm{B}}/R)^4$ was determined in \cite{lt1}.
    The coefficient $c_{10}$ was found in the first order of PT, but its calculation includes an inaccuracy: in the
    charge operator $\hat{Q}$ (see formulas (11)--(13) from
    \cite{lt1}), the $Z$-axis was chosen prior to the
    differentiation. But this axis should be fixed after all differentiations.
    In this last case, all the coefficients $c_{10}^{(1)}$
    become zero, because $c_{10}^{(1)}\sim \langle 0^{\rm{B}}| \hat{Q}_{\rm{B}} |
    0^{\rm{B}}\rangle R^{-1}$ and
    \begin{equation}
     \langle 0^{\rm{B}}| \hat{Q}_{\rm{B}} |
    0^{\rm{B}}\rangle R^{-1} =0.
     \label{2-3d}     \end{equation}
     In particular,
    $\langle 0^{\rm{B}}| \hat{Q}_{\rm{B,q}} |
    0^{\rm{B}}\rangle R^{-1} \sim
    \sum\limits_{\alpha\beta}\delta_{\alpha\beta}\frac{\partial}{\partial
    R_{\alpha}}\frac{\partial}{\partial R_{\beta}}\frac{1}{R} =
    \triangle\frac{1}{R}\equiv 0$, and we get zero only for the
    Coulomb potential. Relation (\ref{2-3d}) is valid, because an helium atom has no moment in the ground state.

      In this connection, we will determine $c_{10}$ in the second order of PT:
     \begin{eqnarray}
     c_{10}^{(2)} &=& -\frac{\langle 00| \hat{U} |00\rangle \langle 10| \hat{U} |00\rangle}{(E_{0,0}^{(0)}-E_{1,0}^{(0)})^2}
      \label{2-4} \\ &+& \sum\limits_{\tilde{m}\neq (\tilde{0},\tilde{0})} \frac{\langle \tilde{m}| \hat{U} |00\rangle \langle 10| \hat{U} |\tilde{m}\rangle}{
     (E_{0,0}^{(0)}-E_{\tilde{m}}^{(0)})(E_{0,0}^{(0)}-E_{1,0}^{(0)})}, \nonumber     \end{eqnarray}
      where  $\tilde{m}=(\tilde{m}_{\rm{A}},\tilde{m}_{\rm{B}})$,      and
   \begin{equation}
     \langle \tilde{m}_{1}| \hat{U} |\tilde{m}_{2}\rangle = \int\Psi^{*\rm{A}}_{\tilde{m}_{1}^{\rm{A}}}\Psi^{*\rm{B}}_{\tilde{m}_{1}^{\rm{B}}}
      \hat{U}
      \Psi^{\rm{A}}_{\tilde{m}_{2}^{\rm{A}}}\Psi^{\rm{B}}_{\tilde{m}_{2}^{\rm{B}}}\rmd\bi{r}_{1}^{\rm{A}}\rmd\bi{r}_{2}^{\rm{A}}\rmd\bi{r}_{1}^{\rm{B}}\rmd\bi{r}_{2}^{\rm{B}}.
          \label{2-5}     \end{equation}
      It follows from  $\hat{U}=\hat{Q}^+_A
      \hat{Q}_{\rm{B}}(1/R)$ \cite{vak}  and  (\ref{2-3d}) that  $\langle 00| \hat{U} |00\rangle =0
      $. Therefore, only the second term on the right-hand side in (\ref{2-4}) is of importance. The main contribution to this term is given
      by the states $\tilde{m}=(1,2)$ and $\tilde{m}=(2,1)$. We have
      \begin{equation}
     c_{10}^{(2)} =\sum\limits_{\tilde{m}\neq (\tilde{0},\tilde{0})}
      \frac{\langle \tilde{m}| \hat{U} |00\rangle \langle 10| \hat{U} |\tilde{m}\rangle}{
     {\sss\triangle}E_{\tilde{m}}{\sss\triangle}E_{1,0}}.
     \label{2-6}     \end{equation}
            As $n_{1}$ and $n_{2}$ increase, the terms in (\ref{2-6}) decrease.
      The calculation indicates that $\bi{d}^{\rm{A}}$ receives a significant contribution from a lot of states at
      $c_{10}$ with $n>2$ (in particular,  $n=5$ and $6$). Since the number of states is huge, it seems impossible to find
      and to sum over all the required states in (\ref{2-6}), as well as all $c_{10}$ with different $n$ and
      $m$.

        Nevertheless, this difficulty can be avoided if the series is summed approximately.
        We note that the difference of energies for an atom $He^4$ is minimum for those of the
        first excited level $1\rm{s}2\rm{p}$ and the
        ground state, ${\sss\triangle}E_{1}(n=2)=21.236$\,eV, and the maximum difference is represented by the ionization energy
        ${\sss\triangle}E_{\rm{ion}}=24.58$\,eV \cite{gerc0}. In
        (\ref{2-6}), a significant contribution is given, for example, by the state
        $1\rm{s}5\rm{p}$ ñ  ${\sss\triangle}E_{1}(n=5)=24.066$\,eV  \cite{gerc0}.
        Thus, the dispersion  of values of ${\sss\triangle}E$ for different excited levels of parahelium
        is approximately $1/7$ of the value of ${\sss\triangle}E$.
        Since ${\sss\triangle}E$ are sufficiently close for different $n$,
        we set them to be equal to a constant  ${\sss\triangle}E$.

 Let us supplement (\ref{2-6}) by the term $\tilde{m}=(\tilde{0},\tilde{0})$
 $\langle 00| \hat{U} |00\rangle \langle 10| \hat{U}
 |00\rangle/(2({\sss\triangle}E)^2)$  (it is equal to zero due to
      $\langle 00| \hat{U} |00\rangle =0 $). With regard for the fact that
      ${\sss\triangle}E_{1,0}\equiv E_{0,0}^{(0)}-E_{1,0}^{(0)} =
      2E_{0}^{(0)}-E_{1}^{(0)}-E_{0}^{(0)}=E_{0}^{(0)}-E_{1}^{(0)}\approx
      -{\sss\triangle}E$ and ${\sss\triangle}E_{\tilde{m}}\equiv
      E_{0,0}^{(0)}-E_{\tilde{m}}^{(0)}=2E_{0}^{(0)}-E_{\tilde{m}_{\rm{A}}}^{(0)}-
      E_{\tilde{m}_{\rm{B}}}^{(0)}\approx -2{\sss\triangle}E$ in (\ref{2-6}),
      we obtain
      \begin{equation}
     c_{10}^{(2)} \approx\frac{1}{2({\sss\triangle}E)^2}\sum\limits_{\tilde{m}}
      \langle 10| \hat{U} |\tilde{m}\rangle \langle \tilde{m}| \hat{U} |00\rangle =
      \frac{1}{2({\sss\triangle}E)^2}  \langle 10| \hat{U^2} |00\rangle,
     \label{2-7}     \end{equation}
  where $ c_{10}^{(2)}$ depends on $n$ and $m$ in the state  $|\tilde{1}\rangle $.
       By writing (\ref{2-7}), we used the well-known property of completeness
       \begin{equation}
     \sum\limits_{\tilde{m}} |\tilde{m}\rangle \langle \tilde{m}| = 1,
     \label{2-8}     \end{equation}
       where the sum is taken over all states of both atoms.
       By considering only the terms with
       $c_{00}$, $c_{10}$, and $c_{01}$ in (\ref{2-1}), relation (\ref{2-3}) yields the first part of the DM of atom A:
       \begin{eqnarray}
  \bi{d}^{\rm{A}}_{1} &=& \int \rmd\bi{r}_1^{\rm{A}} \rmd\bi{r}_2^{\rm{A}} \rmd\bi{r}_1^{\rm{B}}
   \rmd\bi{r}_2^{\rm{B}} \hat{\bi{d}}^{\rm{A}} |c_{00} \Psi^{\rm{A}}_0\Psi^{\rm{B}}_0 \nonumber \\
   &+&  \sum\limits_{\tilde{1}}[c_{10}^{(2)}\Psi^{\rm{A}}_{1}\Psi^{\rm{B}}_0 + c_{01}^{(2)}
  \Psi^{\rm{A}}_0\Psi^{\rm{B}}_{1}]|^{2} \label{2-9}   \\
   &\approx& c_{00}^{*}\sum\limits_{\tilde{1}}c_{10}^{(2)}\langle 00| \hat{\bi{d}}^{\rm{A}}
   |10\rangle + \mbox{c.c.} \nonumber \\
   &=& \frac{c_{00}^{*}}{2({\sss\triangle}E)^2}\sum\limits_{\tilde{1}}\langle 00| \hat{\bi{d}}^{\rm{A}}
   |10\rangle  \langle 10| \hat{U}^2 |00\rangle + \mbox{c.c.}
      \nonumber     \end{eqnarray}
     Here, $\sum\limits_{\tilde{1}}$  is the sum over all states $|\tilde{1}\rangle $, i.e., over $m=0; \pm 1$
       and $n=2, 3, \ldots, \infty$.
         Besides  $|10\rangle$, we include all the remaining states $|\tilde{m}_{\rm{A}}\tilde{m}_{\rm{B}}\rangle $ in (\ref{2-9}):
          $|1^{\rm{A}}\rangle=|1\rm{s}\it{n}\rm{p}\rangle$, and also $|1\rm{s}\it{n}\rm{s}\rangle$,
      $|1\rm{s}\it{n}\rm{d}\rangle$, $|1\rm{s}\it{n}\rm{f}\rangle$, \textit{etc}.; and, in addition to $|0^{\rm{B}}\rangle$, all higher states $|\tilde{m}\rangle$.
      This is possible to make because $\langle 00| \hat{\bi{d}}^{\rm{A}}
   |\tilde{m}_{\rm{A}}\tilde{m}_{\rm{B}}\rangle =0$ for them.
   Then, with regard for (\ref{2-8}), we obtain
     \begin{eqnarray}
  \bi{d}^{\rm{A}}_{1}  &\approx&  \frac{c_{00}^{*}}{2({\sss\triangle}E)^2}\sum\limits_{\tilde{m}_{\rm{A}}\tilde{m}_{\rm{B}}}
  \langle 00| \hat{\bi{d}}^{\rm{A}}
   |\tilde{m}_{\rm{A}}\tilde{m}_{\rm{B}}\rangle  \langle \tilde{m}_{\rm{A}}\tilde{m}_{\rm{B}}| \hat{U^2} |00\rangle + \mbox{c.c.}
   \nonumber \\ &=& \frac{c_{00}}{({\sss\triangle}E)^2} \langle 00|
   \hat{\bi{d}}^{\rm{A}} \hat{U^2}  |00\rangle .
    \label{2-10}     \end{eqnarray}
     Relation (\ref{2-8}) and the normalization conditions for $ \Psi^{\rm{AB}}_0$
     yield
   \begin{equation}
  c_{00}^2 = 1- \sum\limits_{l_{1}j_{1}\neq (0,0)} c_{l_{1}j_{1}}^2 \approx 1 -
  \frac{\langle 00| \hat{U^2} |00\rangle }{4({\sss\triangle}E)^2} = 1+
  \frac{E_{\rm{VdW}}}{2{\sss\triangle}E},
  \label{2-11}     \end{equation}
   where $E_{\rm{VdW}} \approx  - \frac{\langle 00| \hat{U^2} |00\rangle }{2{\sss\triangle}E}
  \approx -2.15 e^{2}(a_{\rm{B}}^{5}/R^6)$ is the van der Waals interaction energy
   for two atoms $He^4$. We have $|E_{\rm{VdW}}| \ll {\sss\triangle}E$;
   therefore, $c_{00}\approx 1$.

     With regard for (\ref{2-1}) and (\ref{2-3}), the total DM of atom A is given by the formula
      \begin{equation}
  \bi{d}^{\rm{A}} =  \sum\limits_{\tilde{m}_{1}^{\rm{A}}\tilde{m}_{2}^{\rm{A}}\tilde{m}_{1}^{\rm{B}}\tilde{m}^{\rm{B}}_{2}}
  c_{\tilde{m}^{\rm{A}}_{1}\tilde{m}^{\rm{B}}_{1}}^{*} c_{\tilde{m}^{\rm{A}}_{2}\tilde{m}^{\rm{B}}_{2}}
   \langle \tilde{m}^{\rm{A}}_{1}\tilde{m}^{\rm{B}}_{1}|\hat{\bi{d}}^{\rm{A}}|\tilde{m}^{\rm{A}}_{2}\tilde{m}^{\rm{B}}_{2}\rangle .
    \label{2-14}     \end{equation}
     The above-derived expression for $\bi{d}_{1}$ is a part of the last relation from the terms
     $c_{00}c_{10}$ in the second order of PT. The formula for $\bi{d}^{\rm{A}}$ includes also the nonzero
contribution in the first order of PT from terms
     of the form $c_{11}^{(1)}c_{21}^{(1)}$,
     $c_{22}^{(1)}c_{21}^{(1)}$, and higher-order ones which behave themselves, respectively, as $\sim (a_{\rm{B}}/R)^7$, $\sim
     (a_{\rm{B}}/R)^9$, \textit{etc}. (see below).
    At $l$ and $j$  which are not equal to zero ($\tilde{0}$) simultaneously,
    the relation
      \begin{equation}
   c_{lj}^{(1)} =\frac{\langle l^{\rm{A}}j^{\rm{B}}| \hat{U} |0^{\rm{A}}0^{\rm{B}}\rangle }{E_0^{(0)} - E_j^{(0)}+
   E_0^{(0)} - E_l^{(0)}}
     \label{2-15}     \end{equation}
     is valid. Since $\langle lj| \hat{U} |00\rangle \neq 0$ only at $l,j \geq 1$, we set $2E_0^{(0)} - E_j^{(0)} - E_l^{(0)}\approx
     -2{\sss\triangle}E$. Since the contribution in the DM from $c_{00}$ is taken into account in
     $\bi{d}_{1}$, we redefine $c_{00}$ in
     (\ref{2-14})  according to (\ref{2-15}) with $
     -2{\sss\triangle}E$ in the denominator (such $c_{00}$ is equal to
     zero). Since $c_{10}^{(1)}=0$, we will write
     also $  -2{\sss\triangle}E$ in the denominator for it instead of $
     -{\sss\triangle}E$. We can now use the
     relation
     (\ref{2-8}), and relation (\ref{2-14}) minus $\bi{d}_{1}$ is reduced to
   \begin{eqnarray}
  \bi{d}^{\rm{A}}_{2}  &\approx & \frac{1}{4({\sss\triangle}E)^2} \sum\limits_{j_{1}l_{1}j_{2}l_{2}}
  \langle 00| \hat{U} |l_{1}j_{1}\rangle  \langle l_{1}j_{1}|
   \hat{\bi{d}}^{\rm{A}} |l_{2}j_{2} \rangle  \langle l_{2}j_{2}|\hat{U} |00\rangle
    \nonumber \\
   &=& \frac{1}{4({\sss\triangle}E)^2} \langle 0^{\rm{A}}0^{\rm{B}}|
   \hat{\bi{d}}^{\rm{A}} \hat{U^2}  |0^{\rm{A}}0^{\rm{B}}\rangle = \frac{\bi{d}^{\rm{A}}_{1}}{4}.
    \label{2-16}     \end{eqnarray}
     Whence we obtain the total DM of atom A in the principal ($\sim   (a_{\rm{B}}/R)^7$) approximation:
    \begin{equation}
  \bi{d}^{\rm{A}} = \bi{d}^{\rm{A}}_{1} + \bi{d}^{\rm{A}}_{2} \approx \frac{5}{4({\sss\triangle}E)^2}
    \langle 0^{\rm{A}}0^{\rm{B}}|
   \hat{\bi{d}}^{\rm{A}} \hat{U^2}  |0^{\rm{A}}0^{\rm{B}}\rangle .
    \label{2-17}     \end{equation}
     Using the formulas for the operator of charge $\hat{Q}$  \cite{lt1}, we represent the
   interatomic potential $\hat{U}$ as the sum
     \begin{equation}
  \hat{U}  = \hat{Q}^+_A \hat{Q}_{\rm{B}}\frac{1}{R} = \hat{U}_{\rm{dd}} +  \hat{U}_{\rm{qd}} + \hat{U}_{\rm{qq}} + \ldots,
    \label{2-18}     \end{equation}
   where
 \begin{equation}
   \hat{U}_{\rm{dd}} = \frac{\hat{\bi{d}}^{\rm{A}} \hat{\bi{d}}^{\rm{B}}}{R^3} -
   \frac{3(\hat{\bi{d}}^{\rm{A}}\bi{R})(\hat{\bi{d}}^{\rm{B}}\bi{R})}{R^5}
    \label{2-19}     \end{equation}
     is the dipole-dipole part of the potential, and
  \begin{eqnarray}
   \hat{U}_{\rm{qd}} &=& \frac{3}{R^4}\sum\limits_{\alpha\beta =1}^{3}\left (\hat{Q}_{\alpha\beta}^{\rm{A}}
   \left [\hat{d}^{\rm{B}}_{\beta}n_{\alpha} + \hat{d}^{\rm{B}}_{\alpha}n_{\beta} -5\hat{\bi{d}}^{\rm{B}} \bi{n}n_{\alpha}n_{\beta} \right ]
    \right )  \nonumber \\
    &+& \frac{3}{R^4}\hat{\bi{d}}^{\rm{B}} \bi{n}
    \sum\limits_{\alpha =1}^{3}\hat{Q}_{\alpha\alpha}^{\rm{A}} -
    (A\leftrightarrow B)
    \label{2-20}     \end{eqnarray}
    is the dipole-quadrupole part.
   The quadrupole-quadrupole part, $\hat{U}_{\rm{qq}}$, includes a negligible correction
   to the DM ($\sim (a_{\rm{B}}/R)^9$ from
   $\hat{U}_{\rm{qd}}\hat{U}_{\rm{qq}}$).  Therefore,
     \begin{equation}
  \hat{U}^2  \approx   \hat{U}_{\rm{dd}}^{2} + 2\hat{U}_{\rm{dd}}\hat{U}_{\rm{qd}} + \hat{U}_{\rm{qd}}^{2}.
    \label{2-21}     \end{equation}
    In view of the parity relative to the inversion, we preserve
    only the term $2\hat{U}_{\rm{dd}}\hat{U}_{\rm{qd}}$. As a result,
    we obtain
   \begin{equation}
  \bi{d}^{\rm{A}} \approx \frac{5}{2({\sss\triangle}E)^2}
    \langle 0^{\rm{A}}0^{\rm{B}}|\hat{\bi{d}}^{\rm{A}} \hat{U}_{\rm{dd}}\hat{U}_{\rm{qd}} |0^{\rm{A}}0^{\rm{B}}\rangle .
    \label{2-22}     \end{equation}
      Let us choose the $Z$-axis along $\bi{R}$: $\bi{R}=R \bi{i}_{z}$.
      Then a nonzero contribution to $\bi{d}^{\rm{A}}$ is given by the terms
   \begin{eqnarray}
   \hat{\bi{d}}^{\rm{A}} \hat{U}_{\rm{dd}}\hat{U}_{\rm{qd}} &=& \frac{6}{R^7}\left \{
   (\hat{d}_{x}^{\rm{B}})^{2}\hat{d}_{x}^{\rm{A}} \hat{d}_{z}^{\rm{A}}\hat{Q}_{xz}^{\rm{A}}
   +  (x\leftrightarrow y) \right. \nonumber \\
    &-& \left. (\hat{d}_{z}^{\rm{B}})^{2}(\hat{d}_{z}^{\rm{A}})^{2}
   (\hat{Q}_{xx}^{\rm{A}} + \hat{Q}_{yy}^{\rm{A}}-2\hat{Q}_{zz}^{\rm{A}}) \right    \}.
    \label{2-23}     \end{eqnarray}
    With regard for the isotropy  $\Psi_{0}=|0\rangle $,  we eventually arrive at
the general formula
            \begin{eqnarray}
  \bi{d}^{\rm{A}} &\approx & \frac{30\bi{i}_{z}}{R^{7}({\sss\triangle}E)^2}
   \langle 0^{\rm{B}}| (\hat{d}_{z}^{\rm{B}})^{2} |0^{\rm{B}}\rangle \left ( \langle 0^{\rm{A}}|\hat{d}_{x}^{\rm{A}} \hat{d}_{z}^{\rm{A}}\hat{Q}_{xz}^{\rm{A}} |0^{\rm{A}}\rangle
   \right. \nonumber \\ &+& \left.
    \langle 0^{\rm{A}}|(\hat{d}_{z}^{\rm{A}})^{2}(\hat{Q}_{zz}^{\rm{A}}-\hat{Q}_{xx}^{\rm{A}}) |0^{\rm{A}}\rangle \right ),
    \label{2-24}     \end{eqnarray}
      which can be rewritten in the following simple form:
    \begin{equation}
  \bi{d}_{\rm{A}} = -D_{7} |e| \frac{a_{\rm{B}}^{8}}{R^7} \bi{n},
  \label{2-25}     \end{equation}
     \begin{equation}
    D_{7}= Z_{\rm{A}}Z_{\rm{B}}\left (\frac{2Ry}{{\sss\triangle}E}\right )^{2}\langle \tilde{r}_{1}^{2}\rangle_{\rm{B}}\langle\tilde{r}_{1}^{4}\rangle_{\rm{A}}.
  \label{2-26}     \end{equation}
   Here,   $\langle \tilde{r}_{1}^{2}\rangle_{\rm{B}}\equiv \langle 0^{\rm{B}}| (r_{1}^{\rm{B}}/a_{\rm{B}})^{2}
   |0^{\rm{B}}\rangle$,   $\langle \tilde{r}_{1}^{4}\rangle_{\rm{A}}\equiv \langle 0^{\rm{A}}| (r_{1}^{\rm{A}}/a_{\rm{B}})^{4}
   |0^{\rm{A}}\rangle$,  $Ry=e^{2}/2a_{\rm{B}}=13.6\,$eV.

The next correction to $\bi{d}_{\rm{A}} $ is of the order of $\sim
      (a_{\rm{B}}/R)^9$, and it is $ \sim (R/a_{\rm{B}})^2$ times
      less than (\ref{2-25}). For He II, $R\sim 3.6\,\mbox{\AA}$, $(R/a_{\rm{B}})^2\simeq 50$; therefore, it can be neglected.
      It is of interest that the separate corrections to $D_{7}$ for
      various $n$ and $m$ in the $(\tilde{1}\tilde{1})$  and $(\tilde{1}\tilde{2})$  states from (\ref{2-1})
       are small, about $0.1$ and less. Moreover, we have an alternating series
      for $\bi{d}_{\rm{A}}$, so that several hundreds of various corrections and about 10 different $n$ significantly contribute to the total DM
      $D_{7}$ (\ref{2-26}).

  The positive sign of $D_{7}$ means that the electron cloud of
  each of two atoms $He^4$ shifts to the neighboring atom. That is, the interatomic space acquires an excess of the
  negative charge.

      To calculate the value of $D_{7}$, we need to know the wave function $|0\rangle $ of the ground state of an atom $He^4$.
      The exact analytic solution is not available for it, but some approximations have been proposed. Within the simplest one-parameter one
        \cite{vak,bete}
    \begin{equation}
    \Psi_0=\varphi_{1\rm{s}}(\bi{r}_1)\varphi_{1\rm{s}}(\bi{r}_2), \
     \varphi_{1\rm{s}}(\bi{r})=\sqrt{Z^{*3}/\pi a_{\rm{B}}^3}\rme^{-rZ^{*}/a_{\rm{B}}}
         \label{2-27}     \end{equation}
         ($Z^* = Z-5/16$),  we get
         $\langle \tilde{r}_{1}^{2}\rangle=3/(Z^*)^2\approx 1.053$,
         $\langle \tilde{r}_{1}^{4}\rangle=45/2(Z^*)^4\approx 2.775$.
         In (\ref{2-26}) for ${\sss\triangle}E$, we take the least (${\sss\triangle}E_{1}(n=2)$) and
     largest (${\sss\triangle}E_{\rm{ion}}$) values, so we obtain $D_{7} \approx 14.32 \mbox{$-$} 19.18$ (here we take into account also
      that $Z_{\rm{A}}=Z_{\rm{B}}=2$).

             The more exact multiparameter $\Psi_0$ were calculated within various methods.
    In \cite{hyl}, it was proposed to seek $\Psi_0$ in the form
        \begin{equation}
    \Psi_{0}(s,t,u)=\varphi(ks,kt,ku),
         \label{2-28}     \end{equation}
         where
    \begin{eqnarray}
    &&\varphi(s,t,u) = A
    \rme^{-s/2}\sum\limits_{l,n,m=0}^{\infty}c_{n,2l,m}s^{n}t^{2l}u^{m}, \label{2-29} \\
    &&s=r_{1}+r_{2}, \ t=r_{1}-r_{2}, \ u=r_{12}. \nonumber
         \end{eqnarray}
                The formalism in the variables $s,u,t$ is given in \cite{zom} in detail.  For the 3-parameter $\Psi_0$  \cite{hyl,zom},
       \begin{eqnarray}
   && \varphi  =A \rme^{-s/2}(1+c_{1}u+c_{2}t^2), \label{2-30} \\
   && c_{1}\approx 0.081, \  c_{2}\approx 0.01, \ k\approx 3.63, \nonumber
              \end{eqnarray}
       the calculation yields
         \begin{eqnarray}
    A &=& (k^{3}/4\pi)\left (4+35c_{1}+48c_{2} \right. \nonumber \\
    &+& \left. 96c_{1}^{2}+576c_{2}^{2}+
    308c_{1}c_{2}\right )^{-1/2}\approx 1.325,
         \label{2-31}     \end{eqnarray}
    \begin{eqnarray}
    \langle \tilde{r}_{1}^{2}\rangle  &=& \frac{768\pi^{2}A^{2}}{k^{8}}\left
    (1+\frac{189}{16}c_{1}+24c_{2} \right. \nonumber \\
    &+& \left. 42c_{1}^{2}+480c_{2}^{2}+  195c_{1}c_{2}\right ),
         \label{2-32}     \end{eqnarray}
     \begin{eqnarray}
    \langle \tilde{r}_{1}^{4}\rangle  &=& \frac{23040\pi^{2}A^{2}}{k^{10}}\left
    (1+\frac{979}{64}c_{1}+52c_{2} \right. \nonumber \\
    &+& \left. 68c_{1}^{2}+1704c_{2}^{2}+
    \frac{8305}{16}c_{1}c_{2}\right ).
         \label{2-33}     \end{eqnarray}
           We have also performed
     the numerical calculations for 6-, 10-, and 20-parameter Hylleraas functions
     \cite{hyl,gerc1,gerc2} and for more general 10-, 39-, and
     80-parameter Kinoshita functions \cite{kin1,kin2} of the form (\ref{2-28}) with
      \begin{equation}
     \varphi(s,t,u) =A
    \rme^{-s/2}\sum\limits_{l,n,m=0}^{\infty}c_{n,m,l}s^{n+1}(t/u)^{2l}(u/s)^{m}
         \label{2-34}     \end{equation}
      which include the negative degrees in the expansion.  The results are given in the
      Table I.
      For the 3-parameter $\Psi_0$, the result is obtained
      numerically and analytically. It is seen from the Table I that the more
      the number of parameters, the less the variation of the results for $\langle\tilde{r}_{1}^{2}\rangle$,
      $\langle \tilde{r}_{1}^{4}\rangle$, and $D_{7}$. Moreover, the
      convergence is quite good. For comparison, we present also
      a deviation of the theoretical ionization potential $J$ of atoms $He^4$ from the
      experimental value, $J^{\rm{exp}}\approx 198310.8\,cm^{-1}$, by the data from \cite{bete,kin2,pek}. It is seen that the
       convergence for the energy is faster, which is natural. Indeed, at the determination of
       $\Psi_0$, namely the energy is minimized. However, the quantities $\langle\tilde{r}_{1}^{2}\rangle$
      and $\langle \tilde{r}_{1}^{4}\rangle$ are determined also
       with a high accuracy. In work \cite{pek},
       $\Psi_0$ was calculated by means of the expansion in Laguerre polynomials.
        Up to 1078 terms were taken into account, and the determined value of $\langle
      \tilde{r}_{1}^{2}\rangle$  (see Table I) coincides with accuracy to 4 decimal points
      with the value obtained by us for the 80-parameter $\Psi_0$.
       This means that the above-determined $\langle\tilde{r}_{1}^{2}\rangle$,
      $\langle \tilde{r}_{1}^{4}\rangle$, and the upper and lower limits of $D_{7}$ for this $\Psi_0$ have error less than $0.1\,\%$.
        The modern more exact methods of calculations of $\Psi_0$ are available
        (see, e.g., \cite{gold,mit}), but the too high accuracy is not required for our task, because it will not increase the accuracy
        of $D_{7}$, since the large error of $D_{7}$ remains unremovable due to the ``smearing'' of ${\sss\triangle}E$.
         \begin{table}
     \caption{Dependence of various quantities on the number
of parameters of $\Psi_0$
     (the wave function of the ground state of $He^4$ atom). }
     \begin{center}
      \begin{tabular}{|c|c|c|c|c|c|c|c|}  \hline
      \mbox{Number of parameters of} $\Psi_0$ & \mbox{Ref.} & $\langle\tilde{r}_{1}^{2}\rangle$ &
      $\langle\tilde{r}_{1}^{4}\rangle$ & $D_{7}^{\rm{min}}$  & $D_{7}^{\rm{max}}$ &  $\bar{D}_{7}$ & $|\frac{J^{\rm{theor}}-J^{\rm{exp}}}{J^{\rm{exp}}}|$  \\ \hline
    $1$  & \cite{bete,vak}  &  $1.053$  & $2.775$ &  $14.32$ & $19.18$ &  $15.53$ &   $5\cdot 10^{-2}$   \\ \hline
    $3$  & \cite{hyl,bete}  &  $1.182$  & $3.815$ &  $22.1$ & $29.6$ &  $23.98$ &  $2.5\cdot 10^{-3}$   \\ \hline
    $6$  & \cite{hyl,bete}  &  $1.188$  & $3.861$ &  $22.47$ & $30.1$ &  $24.38$ &   $5\cdot 10^{-4}$   \\ \hline
    $10$  & \cite{gerc1}  &  $1.249$  & $4.222$ &  $25.82$ & $34.6$ &  $28.02$ &   $1\cdot 10^{-4}$   \\ \hline
   $10$  & \cite{kin1}  &  $1.19$  & $3.91$ &  $22.79$ & $30.54$ &  $24.73$ &   $1\cdot 10^{-6}$   \\ \hline
   $20$  & \cite{gerc2}  &  $1.19326$  & $3.967$ &  $23.19$ & $31.07$ &  $25.16$ &  $2\cdot 10^{-6}$   \\ \hline
   $39$  & \cite{kin2}  &  $1.19314$  & $3.961$ &  $23.15$ & $31.02$ &  $25.12$ &  $1\cdot 10^{-6}$   \\ \hline
   $80$  & \cite{kin2}  &  $1.19346$  & $3.973$ &  $23.23$ & $31.12$ &  $25.2$ &  $2\cdot 10^{-7}$   \\ \hline
   $1078$  & \cite{pek}  &  $1.19348$  & $-$ &  $-$ & $-$ &  $-$ &   $5\cdot 10^{-7}$   \\ \hline
             \end{tabular} \end{center} \end{table}

        Since the main contribution to the DM is given by the states with large $n$,
   for which ${\sss\triangle}E \approx
   {\sss\triangle}E_{\rm{ion}}$, $D_{7}$ should be closer to the lower limit.
   Using formula (\ref{2-26}) and the table of energies of different excited levels \cite{gerc0},
it can be seen that the value of $D_{7}$ belongs to the interval from $(D_{7}^{\rm{max}}+D_{7}^{\rm{min}})/2$
to $D_{7}^{\rm{min}}$. We can estimate the exact value of $D_{7}$, by determining the arithmetic mean for this interval.
   Therefore, we define
    $D_{7}$ as
    \begin{equation}
   \bar{D}_{7} \approx D_{7}^{\rm{min}}+(D_{7}^{\rm{max}}-D_{7}^{\rm{min}})/4.
  \label{2-35}     \end{equation}
    The remaining uncertainty will be taken into account in the error term.
     In view of the result for the 80-parameter $\Psi_0$, we finally have
     \begin{equation}
   D_{7} \approx \bar{D}_{7} \approx 25.2 \pm 2,
  \label{2-36}     \end{equation}
   where the error covers the whole interval
 from $(D_{7}^{\rm{max}}+D_{7}^{\rm{min}})/2$ to $D_{7}^{\rm{min}}$.

   Solution (\ref{2-25}), (\ref{2-36}) is in agreement with (\ref{2-0}), though we have obtained a somewhat larger value for $D_{7}$.
   We note that the formula different from (\ref{2-26}) for $D_{7}$ was derived in \cite{wb1,wb2} with the use of dispersion sums.
   The most exact value, $D_{7}=18.4$, was obtained there on the basis of the 20-parameter $\Psi_0$
   taken from \cite{gerc2}. In this case, certain approximations were used in calculations, but the error arising due to these approximations was not
   evaluated.  For this and more exact functions $\Psi_0$, we have calculated a more larger value of $D_{7}$ (\ref{2-36}).
   The difference of values of $D_{7}$ in (\ref{2-36}) and (\ref{2-0}) is probably related to the approximations made in both calculations.
     We note that the above-proposed calculation is much simpler than that in
   \cite{wb1,wb2}.

      It is possible that $D_{7}$ can be determined more exactly within special numerical methods. But, even in this case, our solution retains its validity,
     since it was obtained by a relatively simple verifiable analytic method. The calculation error for $D_{7}$ is determined by the
approximate character of $\Psi_{0}$ and by the difference of the energies of excited levels of a He atom. However, at a large number of parameters for $\Psi_{0}$,
the first error becomes small (see Table 1), and the second error is approximately equal to the relative dispersion of the values
     of ${\sss\triangle}E$, i.e., to within $15\,\%$. Thus, the method is rather exact. The main error of the method is related to the change of values of ${\sss\triangle}E$ for different levels in sums of the form (\ref{2-6}) by some constant value. Therefore, the method works well only for atoms, for which the energies of excited states are close to one another. One of such atoms is a $He^4$ atom.

 The above-considered (and earlier in \cite{wb1,wb2}) tidal mechanism of the polarization of atoms is, in our opinion,
     a key to the comprehension of electric properties
       of superfluid $He^4$ and other fluids of nonpolar atoms, though we cannot yet exclude
       the contribution of other mechanisms (for example, the
       thermoemf).

    \ack
    The authors thank   A.S. Rybalko, S.I. Shevchenko, G.E. Volovik for the fruitful discussions
              and Yu.V. Shtanov for the valuable remark concerning the calculation.
         The work is performed under the financial support
  of the Division of Physics and Astronomy of the NAS of Ukraine in the frame of a Target program
  of fundamental studies N 0107U000396.

       \end{document}